# A compact vapor-cell optical frequency reference with fractional frequency instability around $10^{-16}$

**Siqi Wu[1, †], Zhenqi Zhang[1, †], Xingyue Liu[1], Chuanshuai Zhu[1], Zhiyuan Wang[1], Zhiyu Ma[1], Hongli Liu[1], Wenhao Yuan[1], Xiaochi Liu[2], Pengfei Wang[2], Feng Zhao[2], Jan Hrabina[3], Jie Zhang[1], Zehuang Lu[1] and Ke Deng[1, *]**

[1] National Gravitation Laboratory (NGL), School of Physics, Huazhong University of Science and Technology, Wuhan 430074, China;

[2] Key Laboratory of Atomic Frequency Standards, Innovation Academy for Precision Measurement Science and Technology, Chinese Academy of Sciences, Wuhan 430071, China;

[3] Institute of Scientific Instruments of the Czech Academy of Sciences, Královopolská 147, 61264 Brno, Czech Republic;

## ABSTRACT

Compact optical frequency reference with high stability is essential for field applications such as navigation and geodesy, yet vapor-cell systems have remained confined to fractional instabilities over $10^{-15}$. Here, we report a molecular iodine reference that reaches an instability of $7 \times 10^{-16}$ at 1000 s and operates at the $10^{-16}$ level from 200 s to 2000 s, surpassing the best reported vapor-cell standards by approximately a factor of three. This achievement is enabled by a monolithic, drift-immune spectroscopic unit bonded to an ultra-low-expansion glass substrate with precision control of key parameters. The entire system occupies only 25 L. The system achieves $5 \times 10^{-15}$ instability at 1 s and reaches the $10^{-16}$ level over the 200–2000 s averaging-time range, representing the first medium-term stability at the $10^{-16}$-level from a compact, field-ready vapor-cell reference. Our work demonstrates that $10^{-16}$ instability can be engineered into portable systems, opening a path to high-precision time-keeping beyond the laboratory.

Precision time-keeping is a cornerstone of modern science and technology. Optical frequency standards based on laser-cooled atoms or ions have reached fractional frequency instabilities below $10^{-18}$, enabling unprecedented tests of fundamental physics [1-5]. Historically, this superior performance came at the cost of substantial size, complexity, and fragility, largely confining such systems to well-controlled laboratory environments. Recent years have witnessed remarkable progress toward transportable and deployable optical clock technologies. The first commercially available optical quantum clock, based on a single trapped $Yb^+$ ion, demonstrated $10^{-17}$-level stability in a two-19-inch-rack form factor [6]. Transportable optical lattice clocks have been shipped across continents and operated within days of arrival, achieving performance comparable to laboratory systems [7]. Despite these advances, a gap remains between the potential and deployability of optical frequency standards for critical applications such as next-generation satellite navigation [8, 9], autonomous time-keeping in field environments [10] and Very Long Baseline Interferometry (VLBI) [11]. Bridging this gap requires robust, compact, and energy-efficient clock architectures that do not compromise core performance metrics.

In parallel, thermal-atomic approaches also offer cavity-competitive stability: a calcium-beam system reaches $10^{-16}$ via Ramsey–Bordé interferometry [12], and Doppler-free three-photon spectroscopy in strontium vapor achieves a ~470 Hz linewidth [13]. Vapor-cell systems, in which atoms or molecules are confined in a simple glass cell, offer a particularly promising route within this class of thermal-ensemble approaches. The molecular iodine reference, an early platform for optical frequency stabilization [14, 15] and length standard [16-21], is undergoing a renaissance. Its low sensitivity to external fields and the mature, frequency-doubled Nd: YAG laser technology [22] that underpins it have driven decades of metrological refinement. Today, this system is finding renewed relevance in space-based missions [23-26] and field-deployable time-keeping [27].

† These authors contributed equally to this work.
* Contact author: ke.deng@hust.edu.cn

Yet, despite decades of refinement, the best iodine cell frequency references have plateaued at fractional frequency instabilities around the $10^{-15}$ level at best [25, 27-29]. This limit arises from inherent physical effects and technical noise sources: homogeneous broadening [30], collisional frequency shifts [31], and technical laser noise—EOM temperature sensitivity, optical power fluctuations, and beam-pointing drift. Consequently, vapor-cell systems have historically been considered less competitive for ultra-stable applications compared to more complex laboratory standards.

The key advance of this work is not the invention of any single component, but rather the system-level integration of established techniques into a compact, field-ready package that achieves $10^{-16}$-level instability in a room-temperature vapor-cell system. Here, we demonstrate that these limitations can be substantially mitigated through a holistic engineering approach that combines drift-immune monolithic integration with active stabilization of key parameters — namely, EOM temperature, optical power, and module temperature. Using the R(56)32-0:$a_1$ line and modulation-transfer spectroscopy (MTS) in a specified saturation pressure (3.3 Pa) cell, we construct an optical frequency reference whose core is a monolithically integrated spectroscopic module: the iodine cell, interference filters and photodetector are permanently epoxy-bonded onto a single ultra-low-expansion glass substrate, effectively suppressing beam pointing drift and achieving exceptional structural stability. This alignment-free core is then housed within a hermetically sealed, precision temperature-stabilized box, providing a first protection against environmental fluctuations. Additionally, key parameters are specifically controlled, including independent high-precision temperature regulation of external electro-optical modulator (EOM) and active stabilization of laser power. Control measurements without ULE bonding, without EOM temperature control, and without optical-power stabilization confirm that the observed $10^{-16}$-level performance relies on the combination of these elements (Supplemental Material).

This design architecture yields a fractional frequency instability in the mid-$10^{-15}$ range at 1 second. More importantly, in the critical 200-second to 2000-second window essential for applications using flywheel local oscillator and coherent signal integration, the instability remains in the $10^{-16}$ level. These results represent the first demonstration of a vapor-cell optical standard attaining stability in the $10^{-16}$ regime in the medium-term stability regime essential for global navigation, VLBI and autonomous field operation. The current system exhibits a long-term drift on timescales beyond $10^4$ s. This drift is attributed to residual effects of the temperature control and laser power control and presents a clear target for future refinement. Our work reestablishes molecular iodine as a leading candidate for next-generation compact optical frequency references.

To achieve frequency discrimination of the hyperfine transition in molecular iodine, we employed modulation transfer spectroscopy and frequency-locked the 532 nm laser output to the R(56)32-0:$a_1$ hyperfine transition line. A schematic of the setup is shown in Fig. 1(a). A Nd:YAG laser emitting at 1064 nm and 532 nm provides the 532 nm output, which is divided by a PBS into pump and probe beams. The pump beam is frequency-shifted by 100 MHz using an acousto-optic modulator (AOM) and subsequently phase-modulated via a resonant (323 kHz) EOM. The EOM is a wedged electro-optic modulator housed within an individual temperature-stabilized enclosure for precise thermal control. The modulation depth is 1.38. The weaker probe beam is shifted by 80 MHz with a separate AOM. Both beams are coupled into polarization-maintaining fibers using fiber couplers and delivered to the iodine spectroscopy module (ISM). After integrating part of the optical path onto the substrate, the whole system only occupies 25 L.

The ISM is shown in the white substrate region in Fig. 1(a), and its internal physical photograph is displayed in Fig. 1(b). The ISM is housed on a ULE glass block substrate with dimensions 33 × 11 × 3 $cm^3$.The mirrors, wave plates and PBS/BS are made of fused silica, and the window surfaces have $\lambda$/10 optical quality. Holders for the wave plates and fiber couplers are made of Invar.

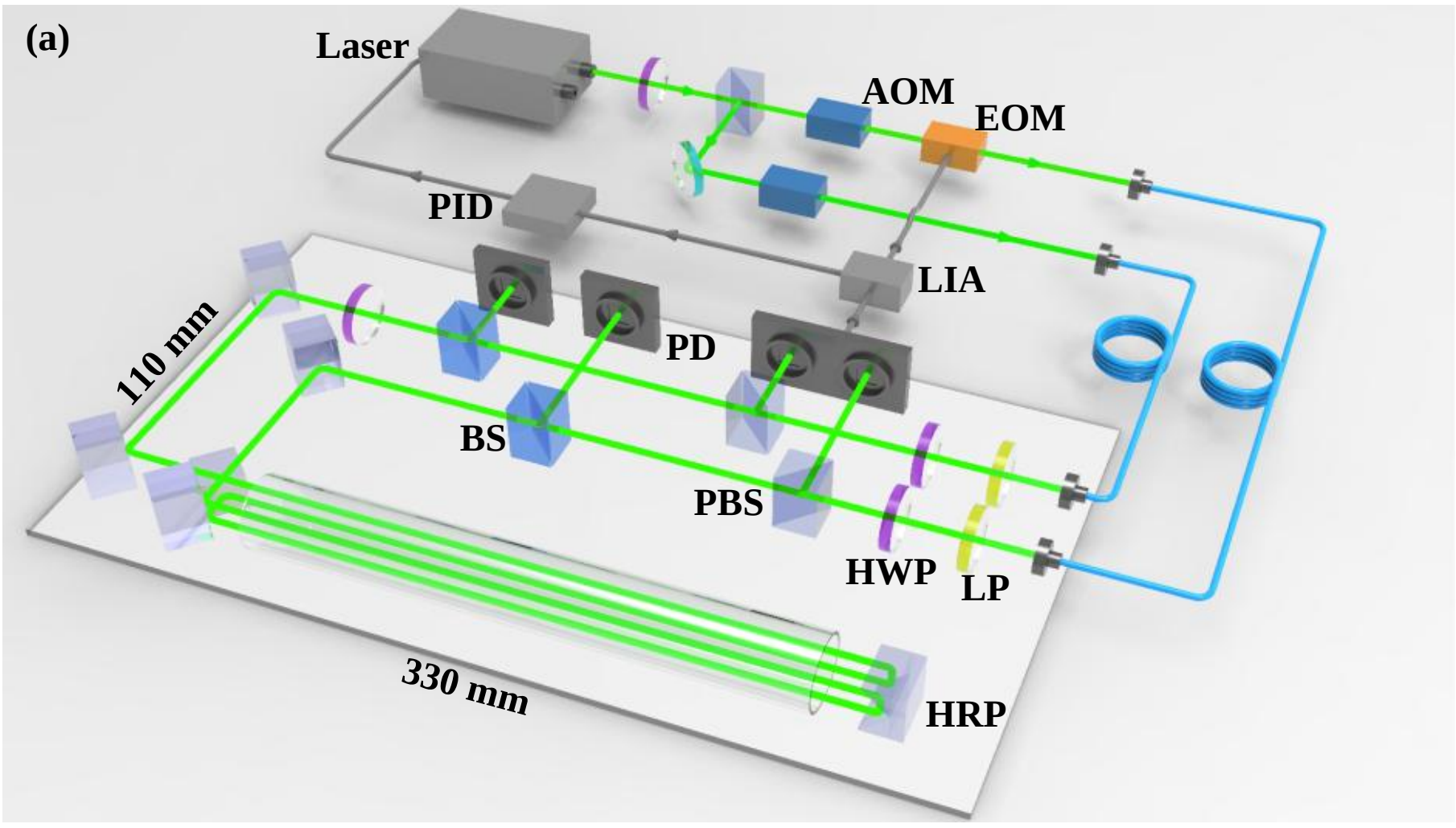

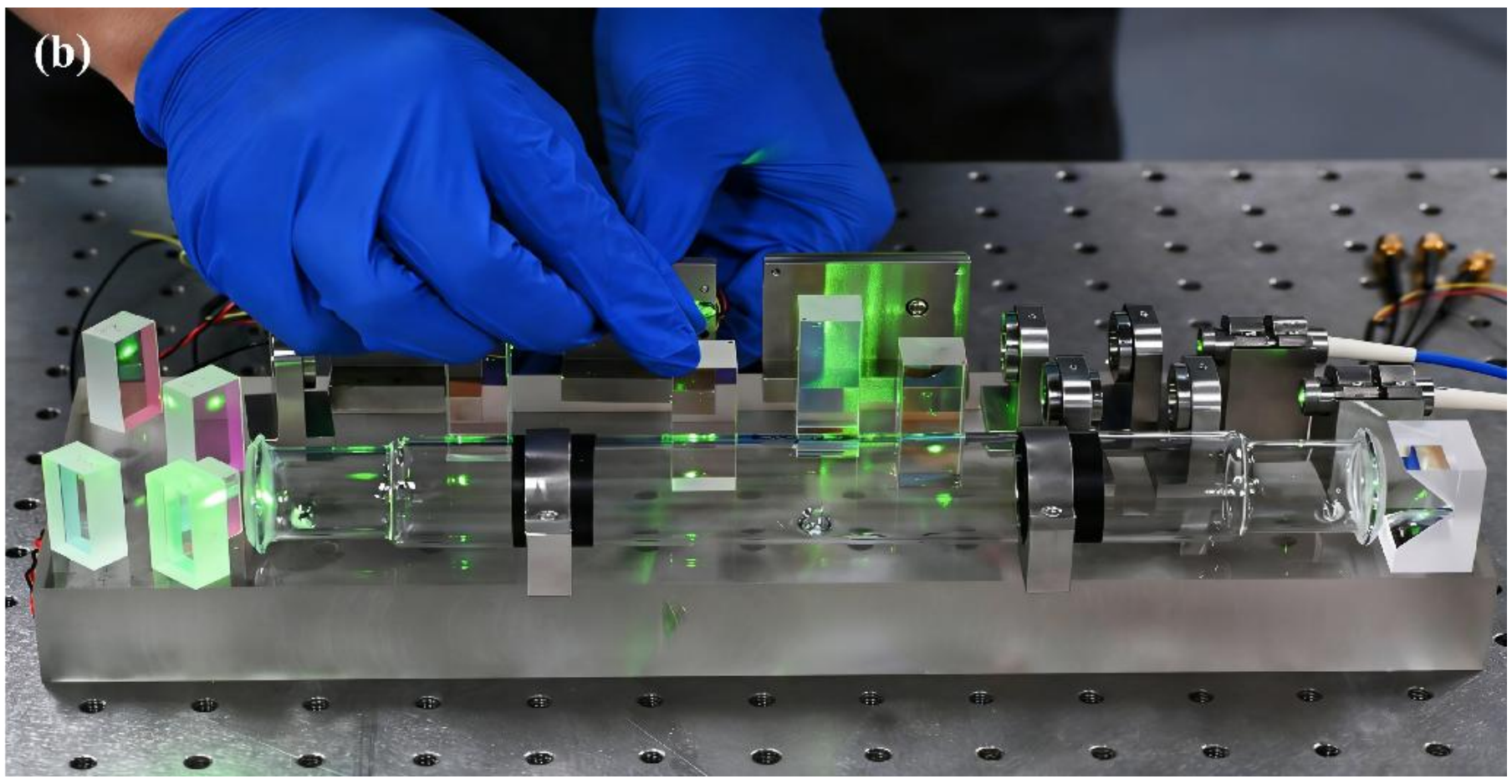

**Figure 1.** (a) Experimental system schematic diagram. HWP: half-wave plate; PBS: polarizing beam splitter; LP: linear polarizer; BS: beam splitter; PD: photo detector; LIA: lock-in amplifier; AOM: acousto-optic modulator; EOM: electro-optic modulator; PID: Proportional-Integral-Derivative; HRP: hollow roof prism. (b) Photograph of the inside of the iodine spectroscopy module.

Inside the ISM, the pump and probe beams are coupled in via two fibres. The probe beam first passes through a half-wave plate and a PBS. A portion of the probe beam is reflected by the PBS to serve as a reference for the balanced photodetector (BPD). Another portion is diverted by a BS for optical power stabilization before being reflected by two mirrors into the iodine cell. Separately, part of the pump beam is likewise sampled by a BS for power stabilization. A half-wave plate is used to ensure that the pump and probe beams maintain orthogonal polarization. Inside the iodine cell, the incident powers are 5.5 mW (pump) and 0.8 mW (probe), with a beam diameter of 2.1 mm for both beams. Within the cell, the two beams are spatially overlapped and propagate in opposite directions to cancel Doppler broadening caused by molecular thermal motion. Upon exiting the cell, the probe beam is reflected by a PBS to the BPD, where the corresponding photocurrent is subtracted from that of the reference light to obtain the MTS signal. All photodetectors are mounted on Invar holders and then epoxy-bonded onto the ULE substrate.

The iodine cell is an unsaturated cell [29] with a specified gas pressure. It is made of fused silica and incorporates wedged optical windows with a 1 ° angle to suppress parasitic back-reflections and etalon effects [32, 33]. Both the inner and outer surfaces of the windows are antireflection-coated for 532 nm wavelength. The cell was fabricated by welding the glass components together. The inner volume was evacuated to a vacuum level of ~ $1 \times 10^{-7}$ Pa, and filled with triple-distilled molecular iodine ($^{127}I_2$) to a final pressure of 3.3 Pa. A four-pass configuration (hollow roof prism at the cell far end) extends the optical path to 1000 mm from a 250 mm physical length.

The error signal from the BPD is demodulated by a lock-in amplifier using quadrature detection to extract the in-phase component. This signal is processed by a servo system based on a PID algorithm for laser frequency stabilization. It is fed back to the laser PZT controller with a servo bandwidth of approximately 13 kHz, and simultaneously to the laser temperature controller with a bandwidth of a few hertz. A custom software routine was developed to automatically identify the R(56)32-0:$a_1$ hyperfine transition from the MTS spectrum and to lock the laser frequency to this line, enabling automated system startup and long-term unattended operation.

The stability of the iodine-stabilized system is characterized by beating the laser's 1064 nm output against an optical frequency comb (OFC) [34, 35], while the comb is locked to an optical frequency reference—specifically, a 30 cm ultra-stable cavity laser. This cavity laser exhibits short-term fractional frequency stability of $1.3 \times 10^{-16}$ at 1 s and $5 \times 10^{-16}$ at 100 s [36]. Due to the frequency drift of the 30 cm cavity, a de-drift process was applied to the raw data. During the stability measurement campaign all the electronics devices are referenced to a hydrogen-maser-based time-scale system [37], which is linked to UTC (National Institute of Metrology, China) via Global Navigation Satellite System (GNSS). The beat signal between the 1064 nm laser output and OFC, which lies in the tens of megahertz range, is filtered, amplified and then downconverted to 1.5 MHz by mixing with a signal generator, followed by amplification, filtering, and input to a frequency counter (Keysight 53230A) for stability analysis.

Owing to mechanical rigidity and active power stabilization, the spectroscopy linewidth approaches the 580 kHz Doppler-broadened limit at this pressure and power. The data presented in Fig. 2 are from a representative run of the measurement campaigns, with a continuous duration of 24,800 s, and the sampling rate was 0.1 s per point. The plotted data correspond to the classical (non-overlapped) Allan deviation. The 1$\sigma$ confidence intervals were evaluated using the Stable32 software package, a community-standard tool for frequency stability analysis. The short-term instability (0.1–1 s) follows approximately $5\times10^{-15}/\tau^{1/2}$, consistent with the theoretical shot-noise limit of $2.9\times10^{-15}/\tau^{1/2}$ (see our earlier work [38] demonstrating ~$3\times10^{-15}/\tau^{1/2}$ at 1 s). This is close to the stability limit for this cell under the designed parameters, which is described in the End Matter. This represents one of the best short-term performances reported for a vapor-cell standard and is currently limited by photodetector shot noise and residual laser frequency/amplitude noise. For integration times beyond 1 s, the instability deviates from the $1/\tau^{1/2}$ scaling due to residual technical noise. The minimum instability of ~$7\times10^{-16}$ at 1000 s represents the technical noise floor of the system after active stabilization with the instability remaining in the $10^{-16}$ regime over the 200-2000 s averaging range. This is the first demonstration of a vapor-cell optical frequency standard attaining stability in the $10^{-16}$ regime in the medium-term regime.

When the averaging time is larger than 2000 s, the frequency stability is slightly worse. This drift is attributed to residual effects within the temperature control and laser power control loop. Nevertheless, after removing the cavity drift, the combined instability at $10^4$ s remains below $3 \times 10^{-15}$. We emphasize that this value represents a corrected upper bound on the iodine reference's long-term performance. The intrinsic drift of the iodine laser was independently measured to be only 12 μHz/s (see End Matter Figure 7), confirming its excellent long-term stability.

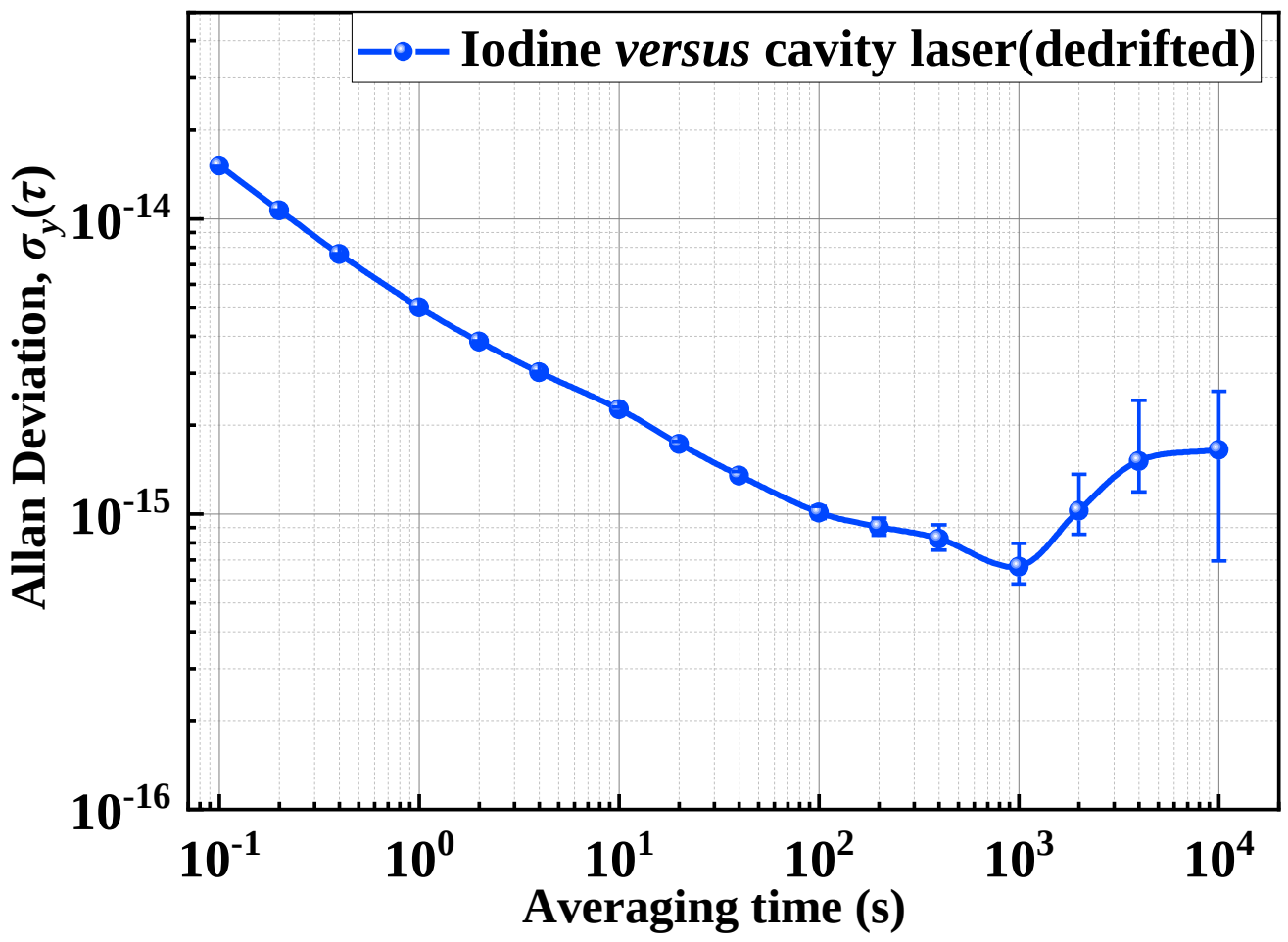

**Figure 2**. Frequency stability of the 532 nm iodine-stabilized laser.

To understand the frequency drift and the long-term instability, the influence of each systematic shift source of the iodine-stabilized laser was investigated. The frequency offsets reported in the following figures correspond to shifts in the iodine spectroscopic lock point as each parameter is varied. These shifts are measured via the heterodyne beat against the cavity laser (referenced through the frequency comb), with the cavity drift independently characterized and de-trended.

The cell-temperature dependence arises primarily from thermally-averaged molecule-molecule collisional shifts, due to the temperature-dependent rovibrational state distribution. We assess this by varying the ISM temperature. In this setup, the iodine cell is bonded inside the ISM, and the entire module is housed within a metal enclosure. The exterior of the housing is wrapped with thermal insulation to provide passive temperature control, ensuring that temperature fluctuations inside the ISM remain below 10 mK. By monitoring the frequency change as the temperature within the ISM is increased, the temperature-induced frequency shift coefficient was measured to be 146.5(2) Hz/K, as shown in Fig. 3(a), corresponding to a frequency instability of 1.47 Hz. The pressure/collisional shift is not treated as a separate contribution, as the iodine cell is unsaturated and the vapor pressure is determined by the cell temperature; both pressure and collisional shifts are linearly correlated with temperature, and their contribution is therefore captured by the cell temperature instability entry in Table 1. While no precise BBR value exists for this transition, any temperature-dependent BBR contribution within our narrow temperature control range is inherently included in our measured temperature sensitivity and thus covered by the temperature instability budget.

Residual amplitude modulation (RAM) noise originates primarily from non-ideal effects such as polarization-related influences and birefringence in the electro-optic phase modulator, as well as parasitic interference within the optical path. The wedged EOM remains slightly temperature-sensitive. The EOM temperature was varied over a 1 K range (293.9–294.9 K) to determine the local frequency shift coefficient at the operating setpoint. In practice, the EOM temperature is stabilized to better than 0.1 K. Over a broader range, the shift exhibits a sinusoidal dependence due to RAM-induced frequency shifts (see Ref. [28], Eq. (10)); the chosen setpoint corresponds to a measured coefficient of approximately 65 Hz/K, which is the minimum accessible value.

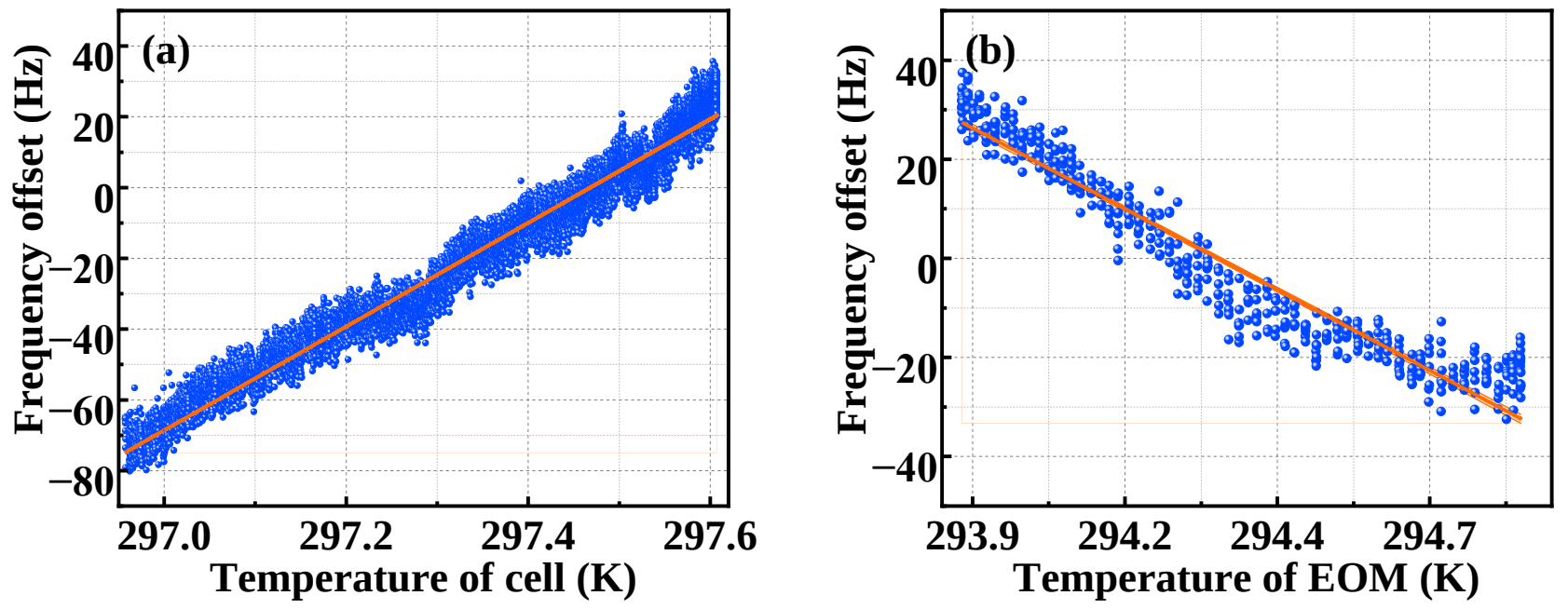

**Figure 3.** Relationship between the measured frequency offset and (a) temperature of the cell, (b) temperature of EOM. Orange lines indicate the linear fit to the data points.

In an alternating electromagnetic field, the rapid electric field oscillation between positive and negative amplitudes suppresses the first-order Stark effect in systems lacking a permanent electric dipole moment. However, this oscillation concurrently enhances the second-order Stark effect, ultimately resulting in shifts of the energy levels. The R(56)32-0 transition line of iodine molecules corresponds to a magnetically insensitive state. The influence of the optical field on this transition is primarily manifested through the AC Stark effect induced by the optical electric field, which gives rise to spectral line frequency shifts, also referred to as power-induced frequency shifts.

By adjusting the optical power of the pump and probe beams, the power-induced frequency shift coefficients were measured to be 330(6) Hz/mW for the pump beam and 1138(6) Hz/mW for the probe beam, as shown in Fig. 4. In the experiment, the pump beam power entering the cell was 5.5 mW. After power stabilization, the power instability remained below 7 μW, resulting in a corresponding frequency instability of 2.31 Hz. The probe beam power entering the cell was 0.8 mW and the power instability is 0.6 μW, corresponding to a frequency instability of 0.68 Hz.

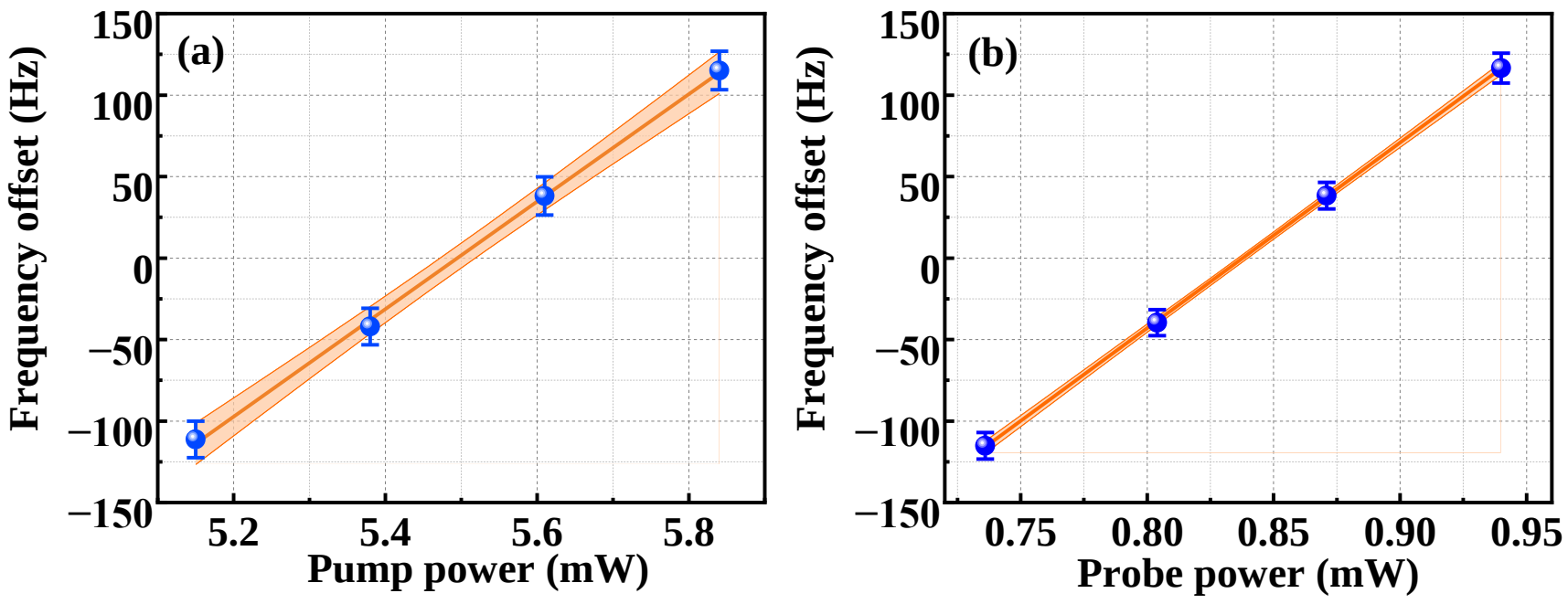

**Figure 4.** Relationship between the measured frequency offset and (a) pump power, (b) probe power. Orange lines indicate the linear fit to the data points.

In laser frequency stabilization based on the modulation transfer spectroscopy, the modulation parameter, including the modulation frequency, modulation depth and the demodulation phase will all influence the locked laser frequency. Varying the EOM modulation frequency yields 61(1) Hz/kHz, as shown in Fig. 5(a), corresponding to 0.06 Hz instability. Since the frequency discrimination signal is obtained by comparing the different absorption intensities of the left and right sidebands of phase-modulated light in a nonlinear absorption medium, the selection of modulation depth is crucial. By varying the modulation voltage, the frequency shift due to modulation depth was measured to be −306(15) Hz, as shown in Fig. 5(b), with an instability of 0.31 Hz. The demodulation phase shift coefficient was determined to be −8.6(2) Hz/degree by adjusting the phase of the demodulation signal, as illustrated in Fig. 5(c). Since the lock-in amplifier employs quadrature demodulation, its associated frequency shift is considered negligible. The demodulation phase is determined not only by the programmed lock-in phase but also by any temperature-dependent electronic phase shifts introduced by the detector, cables, and connectors. We performed a measurement of the phase stability of the entire electronic chain. For a long-term measurement of 24 hours, we observed a phase fluctuation of approximately 0.005 degree, corresponding to a frequency instability of 0.04 Hz, as detailed in Supplemental Material Figure 9.

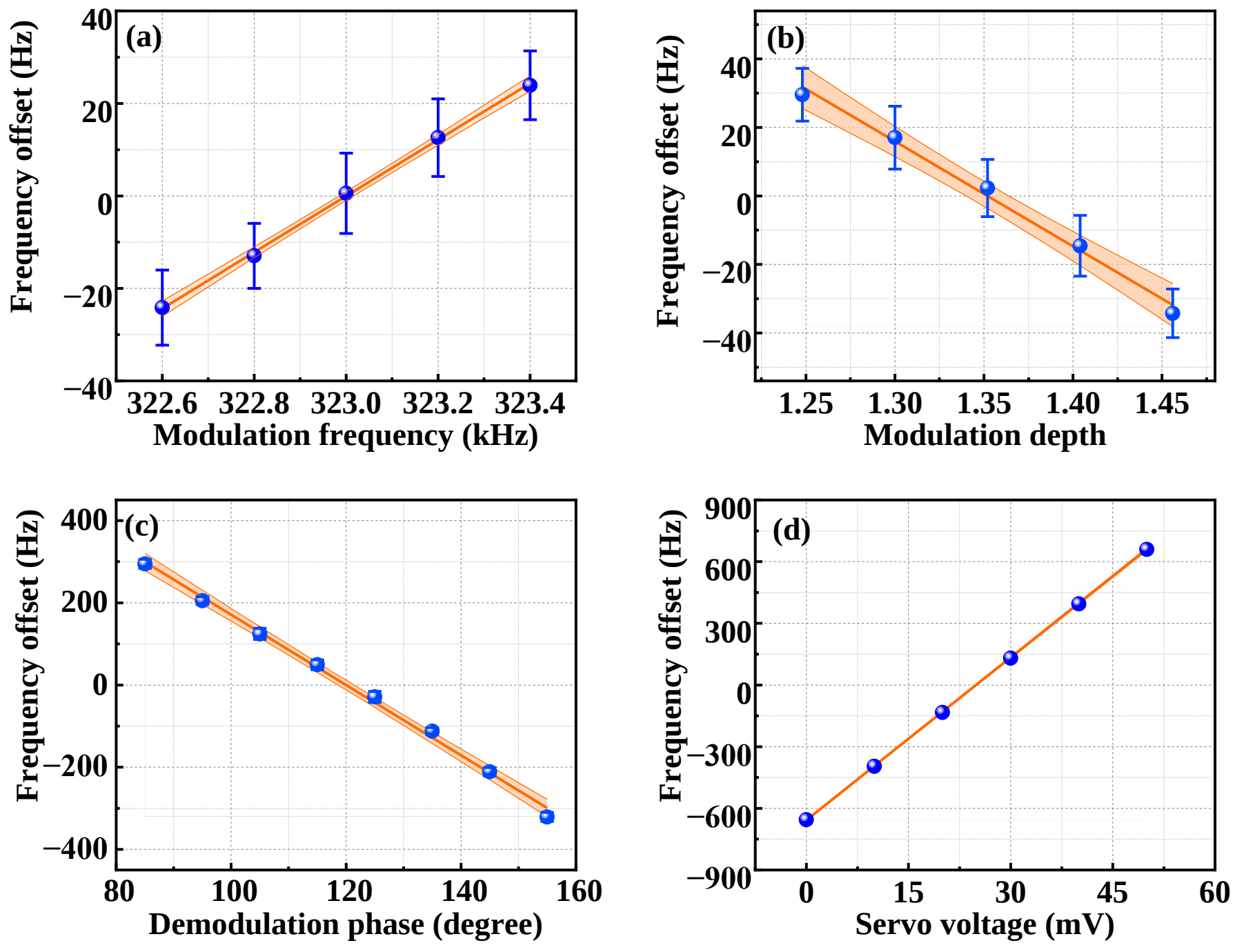

**Figure 5.** Relationship between the measured frequency offset and (a) modulation frequency, (b) modulation depth, (c) demodulation phase, and (d) servo voltage. Orange lines indicate the linear fit to the data points.

In the experiments, the frequency at the zero-crossing point of the error signal is typically selected as the frequency discrimination point. The frequency discrimination slope of this servo system is measured to be 26.40(4) Hz/mV, as shown in Fig. 5(d). Given that voltage offset fluctuations consistently remain below 10 μV, the corresponding frequency instability attributable to electronic servo noise amounts to approximately 0.26 Hz.

For the second-order Zeeman shift, we measured the residual magnetic field near the spectroscopy module to be $B_0$ = 0.406 G. Using the quadratic Zeeman coefficient $C$ = 0.26 Hz/G$^2$ [39], we calculate a fractional frequency shift of 0.04 Hz, which is included as an upper bound in Table 1. The iodine laser and the optical frequency comb were placed on the same optical table, so only a 2-meter-long fiber was used to deliver the 1064-nm laser to the comb for beat-note detection. For a 2 m length fiber, its contribution is measured to be below $2\times10^{-16}/\sqrt{\tau}$ level in fractional frequency, completely negligible for the reported instability. We have calculated the line-pulling contributions from the 14 nearest neighboring hyperfine components. The total line-pulling shift is estimated to be -3.15 kHz. Since this is a static shift that does not fluctuate with time, we do not include them in Table 1.

The most significant contributions to the frequency instability of the compact iodine-stabilized laser are summarized in Table 1. The primary sources of noise are the iodine cell temperature, EOM temperature, and optical power fluctuations. The total frequency instability is 2.92 Hz (relatively $5.2\times10^{-15}$), which is an expected upper bound on the long-term instability.

**Table 1.** Most significant contributions to the instability of the iodine-stabilized laser

| Source | Sensitivity | Parameter instability | Frequency instability (Hz) |
|---|---|---|---|
| Temperature of cell | 146.5(2) Hz/K | 10 mK | 1.47 |
| Temperature of EOM | −65.2(8) Hz/K | 10 mK | 0.65 |
| Pump power | 330(6) Hz/mW | 7 μW | 2.31 |
| Probe power | 1138(6) Hz/mW | 0.6 μW | 0.68 |
| Modulation frequency | 61(1) Hz/kHz | 1 Hz | 0.06 |
| Modulation depth | 306(15) Hz | 1 mHz | 0.31 |
| Demodulation phase | −8.6(2) Hz/degree | 0.005 degree | 0.04 |
| Servo electronics | 26.40(4) Hz/mV | 10 μV | 0.26 |
| Magnetic-field sensitivity | 0.26 Hz/$G^2$ | 0.4 G | 0.04 |
| **Total** | | | **2.92** |
| **Relative instability** | | | **$5.2\times10^{-15}$** |

Our work overturns the notion that vapor-cell standards are intrinsically limited to the $10^{-14}$ to $10^{-15}$ range. By demonstrating medium-term (minute to hour) stability at the $10^{-16}$ level with an iodine-stabilized laser-a first for any vapor-cell optical standard-we push the performance ceiling for compact optical frequency reference. By combining a passively stable, drift-immune monolithic spectroscopic core with precision control of critical physical parameters, we suppress the technical noise that has historically limited compact vapor-cell references to the $10^{-15}$ level. The resulting system achieves a minimum instability of approximately $7 \times 10^{-16}$ at 1000 s, representing the technical noise floor of our integrated system. This result proves that the classical barriers of homogeneous broadening and collisional shifts in room-temperature vapors are not ultimate boundaries; with disciplined system engineering, vapor-cell standards can access performance domains once exclusive to the most advanced cold-atom platforms.

The attainment of $10^{-16}$ medium-term stability unlocks new applications for compact optical standards. These devices can now function as ultra-stable flywheel oscillators to maintain phase coherence across geographically separated sites, such as in next-generation GNSS or very-long-baseline interferometry networks. Furthermore, they enable high-resolution geopotential measurements on mobile platforms and can serve as synchronized time sources for emerging distributed quantum sensing networks. By pushing beyond the $10^{-15}$ level, we move beyond merely miniaturizing laboratory instruments and towards embedding fundamental metrological performance into the fabric of field-based science and technology. This work thus repositions vapor-cell optical standards as cornerstone technologies for future advancements in precision timing, navigation, and fundamental physics beyond the laboratory walls.

## Acknowledgments.

The authors gratefully thank Jingbiao Chen and Duo Pan for their valuable advice. This study is funded by National Key Research and Development Program of China (2022YFB3904000, 2022YFC2204002) and Strategic Priority Research Program of the Chinese Academy of Sciences (XDA0520503).

## Data availability.

Raw data that support the findings of this study are available from the corresponding author upon reasonable request.

## References

[1] W. F. McGrew et al., Nature **564**, 87 (2018).

[2] E. Oelker et al., Nature Photon. **13**, 714 (2019).

[3] B. Zhang, et al., Phys. Rev. Lett. **136**, 053202 (2026).

[4] A. P. Hilton et al., Nature Commun. **16**, 6063 (2025).

[5] M. C. Marshall et al., Phys. Rev. Lett. **135**, 033201 (2025).

[6] TOPTICA Photonics, "TOPTICLOCK Optical Quantum Clock," https://www.toptica.com/products/optical-quantum-clocks/topticlock.

[7] T. Bothwell et al., Opt. Lett. **50**, 646 (2025).

[8] T. Schuldt et al., GPS Solut. **25**, 83 (2021).

[9] T. Wegehaupt et al., J. Phys. Conf. Ser. **2889**, 012012 (2024).

[10] K. Döringshoff, T. Schuldt, E. V. Kovalchuk, J. Stühler, C. Braxmaier, and A. Peters, Appl. Phys. B **123**, 183 (2017).

[11] Y. Asaki, B. A. Pampliega, P. G. Edwards, S. Iguchi, and E. J. Murphy, Nat. Rev. Methods Primers **3**, 89 (2023).

[12] J. Olson et al., Phys. Rev. Lett. **123,** 073202 (2019).

[13] G. Panelli et al., Phys. Rev. A **111**, 033112 (2025).

[14] M. L. Eickhoff and J. L. Hall, IEEE Trans. Instrum. Meas. **44**, 155 (1995).

[15] J. Ye, L. S. Ma, and J. L. Hall, Phys. Rev. Lett. **87**, 270801 (2001).

[16] F. Riehle, P. Gill, F. Arias, and L. Robertsson, Metrologia **55**(2), 188 (2018).

[17] F. L. Hong, J. Ishikawa, Y. Zhang, and R. Guo, Opt. Commun. **235(4)**, 377 (2004).

[18] E. J. Zang, J. P. Cao, Y. Li, C. Y. Li, Y. K. Deng, and C. Q. Gao, IEEE Trans. Instrum. Meas. **56**, 673 (2007).

[19] J. Lazar, J. Hrabina, P. Jedlicka, and O. Cip, Metrologia **46,** 450 (2009).

[20] K. Nyholm, M. Merimaa, T. Ahola, and A. Lassila, IEEE Trans. Instrum. Meas. **52**(2), 284 (2003).

[21] F. H. Cheng, N. Jin, F. L. Zhang, H. Li, Y. B. Du, J. Zhang, K. Deng, and Z. H. Lu, Chin. Phys. B **30**, 050603 (2021).

[22] J. R. Zhang, X. P. Wu, Z. L. Deng, D. Y. Zhu, C. K. Liu, C. L. Guo, J. Zhang, Z. H. Lu, and W. H. Yuan, Chin. J. Lasers **53**(02), 0201004 (2026).

[23] F. Kuschewski et al., GPS Solutions **28**, 10 (2024).

[24] K. Döringshoff et al., Phys. Rev. Appl. **11**, 054068 (2019).

[25] T. Schuldt, K. Döringshoff, E. V. Kovalchuk, A. Keetman, J. Pahl, A. Peters, and C. Braxmaier, Appl. Opt. **56**, 1101 (2017).

[26] T. Schuldt, K. Döringshoff, M. Oswald, E. V. Kovalchuk, A. Peters, and C. Braxmaier, Int. J Mod. Phys. D **28**, 1845002 (2019).

[27] J. D. Roslund et al., Nature **628**(8009), 736-740 (2024).

[28] Z. Q. Zhang, Z. Y. Wang, H. L. Liu, W. H. Yuan, W. You, J. Zhang, K. Deng, and Z. H. Lu, Class. Quantum. Grav. **40**, 225001 (2023).

[29] Z. Q. Zhang, W. You, X. Y. Liu, H. L. Liu, W. H. Yuan, J. Zhang, K. Deng, and Z. H. Lu, Rev. Sci. Instrum. **95**, 063003 (2024).

[30] R. R. Gamache and B. Vispoel, J. Quant. Spectrosc. Radiat. Transfer. **217**, 440 (2018).

[31] S. Y. Chen and M. Takeo, Rev. Mod. Phys. **29**, 20 (1957).

[32] J. Oulehla, P. Pokorný, J. Hrabina, M. Holá, O. Číp, and J. Lazar, Opt. Express **27**(7), 9361 (2019).

[33] J. Hrabina, M. Šarbort, O. Acef, F. D. Burck, N. Chiodo, M. Holá, O. Číp, and J. Lazar, Appl. Opt. **53**, 7435 (2014).

[34] S. T. Cundiff and J. Ye, Rev. Mod. Phys. **75**, 325 (2003).

[35] S. A. Diddams, K. Vahala, and T. Udem, Science **369**, eaay3676 (2020).

[36] Z. Y. Wang, Z. Y. Ma, W. Z. Wei, J. L. Chang, J. X. Zhang, Q. Y. Wu, W. H. Yuan, K. Deng, Z. H. Lu, and J. Zhang, Rev. Sci. Instrum. **95**, 053002 (2024).

[37] Z. Y. Ma et al., Metrologia, **62**, 045002 (2025).

[38] S. Wu, X. Liu, Z. Zhang, C. Zhu, Q. Wu, J. Zhang, Z. Wang, W. Yuan, H. Liu, J. Zhang, K. Deng, and Z. Lu, Chin. J. Lasers **53**, 1101016 (2026).

[39] J. Roslund, "Molecular iodine optical clocks for the fields," presented at the 2024 European Frequency and Time Forum, Neuchâtel, Switzerland, 2024.

[40] V. Schkolnik et al., EPJ Quantum Technol. **4**, 9 (2017).

[41] L. S. Ma, L. E. Ding, and Z. Y. Bi, Appl. Phys. B **51**, 233 (1990).

[42] C. Ishibashi, J. Ye, and J. L. Hall, Proc. SPIE 4634, Methods for Ultrasensitive Detection II, (28 March 2002).

## End Matter

*Appendix A: Short-term stability limit analysis*—For an iodine molecular frequency standard system that uses atomic/molecular transition spectral lines as the frequency reference, the frequency stability of the system limited by shot noise can be expressed as [40] [41]:

$$\sigma = \frac{\Gamma}{v}\frac{1}{SNR}\frac{1}{\sqrt{\tau}}, \tag{1}$$

where $\Gamma$ is the total linewidth of the reference spectral line and $v$ is the laser frequency. $\Gamma$ can be expressed as [40]:

$$\Gamma = \frac{\left(\sqrt{1+S_{Probe}}+\sqrt{1+S_{Pump}}\right)}{2}\left(\Gamma_{na}+\Gamma_{tof}+\Gamma_{pr}\right), \tag{2}$$

where $S_{pump}$ and $S_{probe}$ are the saturation parameters corresponding to the pump light and the probe light in the system. $\Gamma_{na}$ is the natural linewidth of the spectral line (220 kHz for this line), $\Gamma_{tof}$ is the time-of-flight broadening of the spectral line (22 kHz here), and $\Gamma_{pr}$ is the pressure broadening of the spectral line. A pressure broadening coefficient (74 kHz/Pa) is given in [42].

The *SNR* in Equation (1) is the signal-to-noise ratio of the system's error signal at a filter bandwidth of 1 Hz. It can be expressed as follows:

$$SNR = \frac{i_{sig}}{i_t + i_s + i_R + i_i}, \tag{3}$$

where $i_{sig}$ is the signal intensity of the system's error signal, $i_t$ is the thermal noise introduced by the photodetector, $i_s$ is the shot noise of the system, $i_R$ is the residual amplitude modulation noise, and $i_i$ is the laser intensity noise. Here we only consider the shot noise of the system.

$$i_s = \sqrt{\frac{2\eta e^2 fI}{hv}}. \tag{4}$$

$$i_{sig} = \frac{J_0(\beta)J_1(\beta)}{2}\frac{\Delta\alpha\, pLI\eta e}{hv}. \tag{5}$$

Here $I$ is the laser intensity, $\eta$ is the quantum efficiency of the detector, $J_n(\beta)$ is $n$-th order Bessel function, $f$ is the filter bandwidth, $\Delta\alpha$ is the saturation signal contrast, $L$ is the effective absorption path length of the cell. Based on the experimental parameters summarized in Table 2, the theoretical limit of the system's frequency stability can be calculated as $2.9\times10^{-15}/\tau^{1/2}$. The curve in Fig. 2 shows a stability of $4.6\times10^{-15}/\tau^{1/2}$ between 0.1 s~ 1s. This difference is caused by the optical path alignment not being optimal during the module bonding process, as there is no spectral signal to monitor during the bonding process.

**Table 2.** Summary of experimental parameters

| Experimental parameters | Symbol | Value |
|---|---|---|
| Modulation frequency | $\Omega$ | 323 kHz |
| Modulation depth | $\beta$ | 1.38 |
| Pump power | $I_{pump}$ | 5.5 mW |
| Probe power | $I_{probe}$ | 0.8 mW |
| Beam diameter | $d$ | 2.1 mm |
| Iodine cell pressure | $p$ | 3.3 Pa |
| Effective absorption path length | $L$ | 1 m |
| Quantum detector efficiency | $\eta$ | 0.25 A/W |

*Appendix B: Drift measurements*—The de-drift procedure was applied by subtracting a linear drift of 4.06 mHz/s from the total beat-note data. This drift rate was determined from the overall slope of the beat frequency between the iodine-locked laser and the cavity laser over the full measurement duration (Fig. 6). The total beat drift (4.06 mHz/s) represents the difference between the iodine drift and the cavity drift.

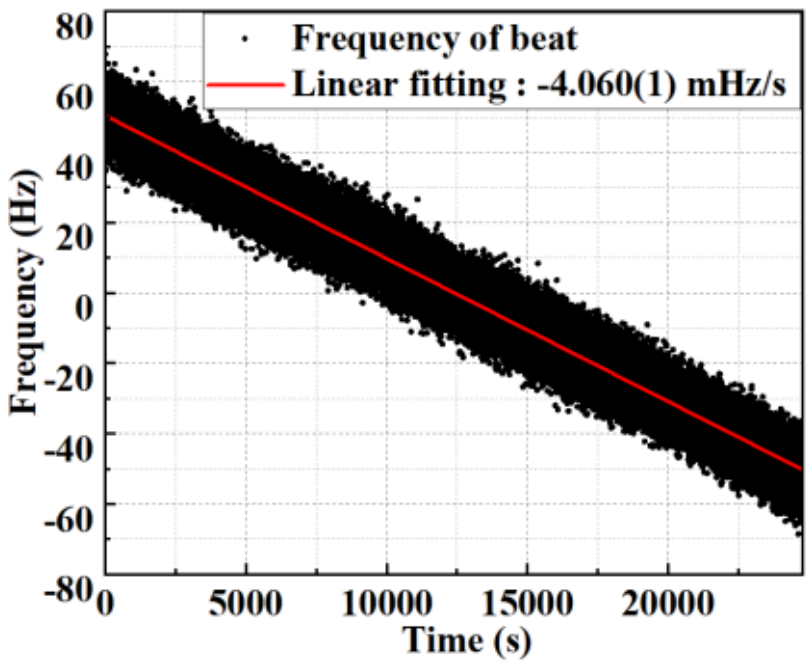

Figure 6. Raw frequency drift of the beat between the iodine-locked laser and the cavity laser, measured over 24,800 s (black), with a linear fit (red). The fitted slope is 4.06 mHz/s.

We have also separately characterized the drifts of both the iodine reference and the cavity using the GNSS-referenced hydrogen maser. The iodine laser drift was measured over four days and found to be approximately 12.0(8) μHz/s (Fig. 7).

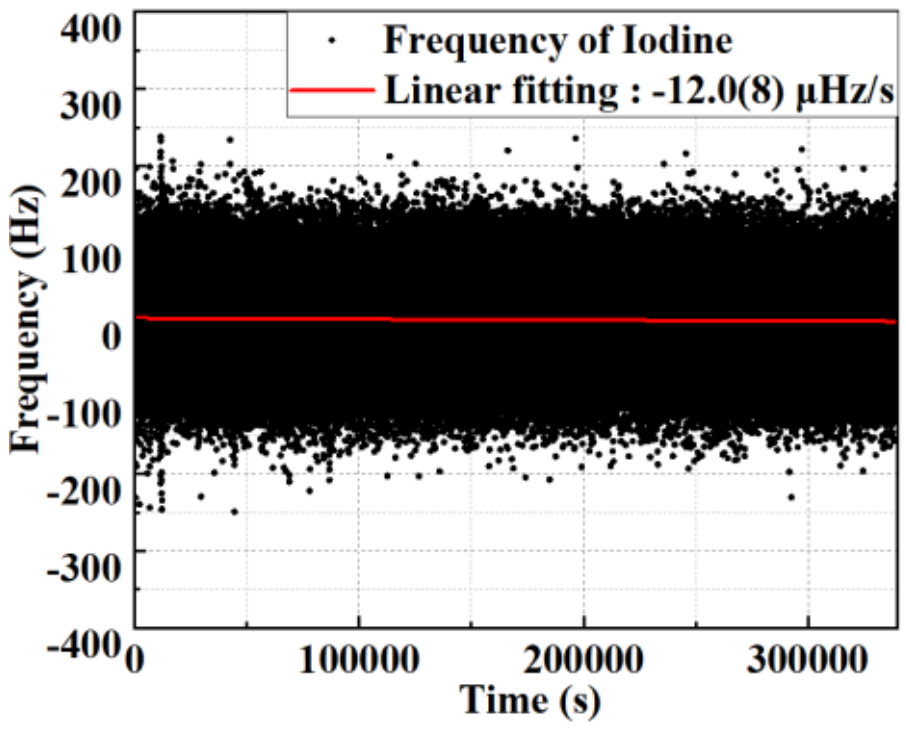

Figure 7. Raw frequency drift of the iodine laser measured over four days (black), with a linear fit (red). The fitted slope is 12.0(8) μHz/s.